\documentclass[aps,prl,reprint,superscriptaddress,amssymb]{revtex4-1}
\usepackage{graphicx}				
\usepackage{color,amsmath}
\usepackage[normalem]{ulem}

\begin{document}


\title{Proximity Effect Transfer from NbTi into a \\ Semiconductor Heterostructure via Epitaxial Aluminum }



\author{A.~C.~C.~Drachmann}%
\affiliation{Center for Quantum Devices and Station Q Copenhagen, Niels Bohr Institute\\
	University of Copenhagen, Universitetsparken 5, 2100 Copenhagen, Denmark}

\author{H.~J.~Suominen}%
\affiliation{Center for Quantum Devices and Station Q Copenhagen, Niels Bohr Institute\\
	University of Copenhagen, Universitetsparken 5, 2100 Copenhagen, Denmark}

\author{M.~Kjaergaard}%
\affiliation{Center for Quantum Devices and Station Q Copenhagen, Niels Bohr Institute\\
	University of Copenhagen, Universitetsparken 5, 2100 Copenhagen, Denmark}
	
\author{B.~Shojaei}%
\affiliation{Materials Department, University of California, Santa Barbara, CA 93106, USA}

\author{C.~J.~Palmstr\o m}%
\affiliation{Materials Department, University of California, Santa Barbara, CA 93106, USA}
\affiliation{Department of Electrical and Computer Engineering, University of California, Santa Barbara, CA 93106, USA}

\author{C.~M.~Marcus}%
\affiliation{Center for Quantum Devices and Station Q Copenhagen, Niels Bohr Institute\\
	University of Copenhagen, Universitetsparken 5, 2100 Copenhagen, Denmark}

\author{F.~Nichele}%
\affiliation{Center for Quantum Devices and Station Q Copenhagen, Niels Bohr Institute\\
	University of Copenhagen, Universitetsparken 5, 2100 Copenhagen, Denmark}


\date{\today}

\begin{abstract}
We demonstrate the transfer of the superconducting properties of NbTi---a large-gap high-critical-field superconductor---into an InAs heterostructure via a thin intermediate layer of epitaxial Al. Two device geometries, a Josephson junction and a gate-defined quantum point contact, are used to characterize interface transparency and the two-step proximity effect. In the Josephson junction, multiple Andreev reflection reveal near-unity transparency, with an induced gap $\Delta^*=0.50~\mathrm{meV}$ and a critical temperature of $7.8~\mathrm{K}$. Tunneling spectroscopy yields a hard induced gap in the InAs adjacent to the superconductor of $\Delta^*=0.43~\mathrm{meV}$ with substructure characteristic of both Al and NbTi.
\end{abstract}


\maketitle


\begin{figure*}
\includegraphics[width=0.88\textwidth]{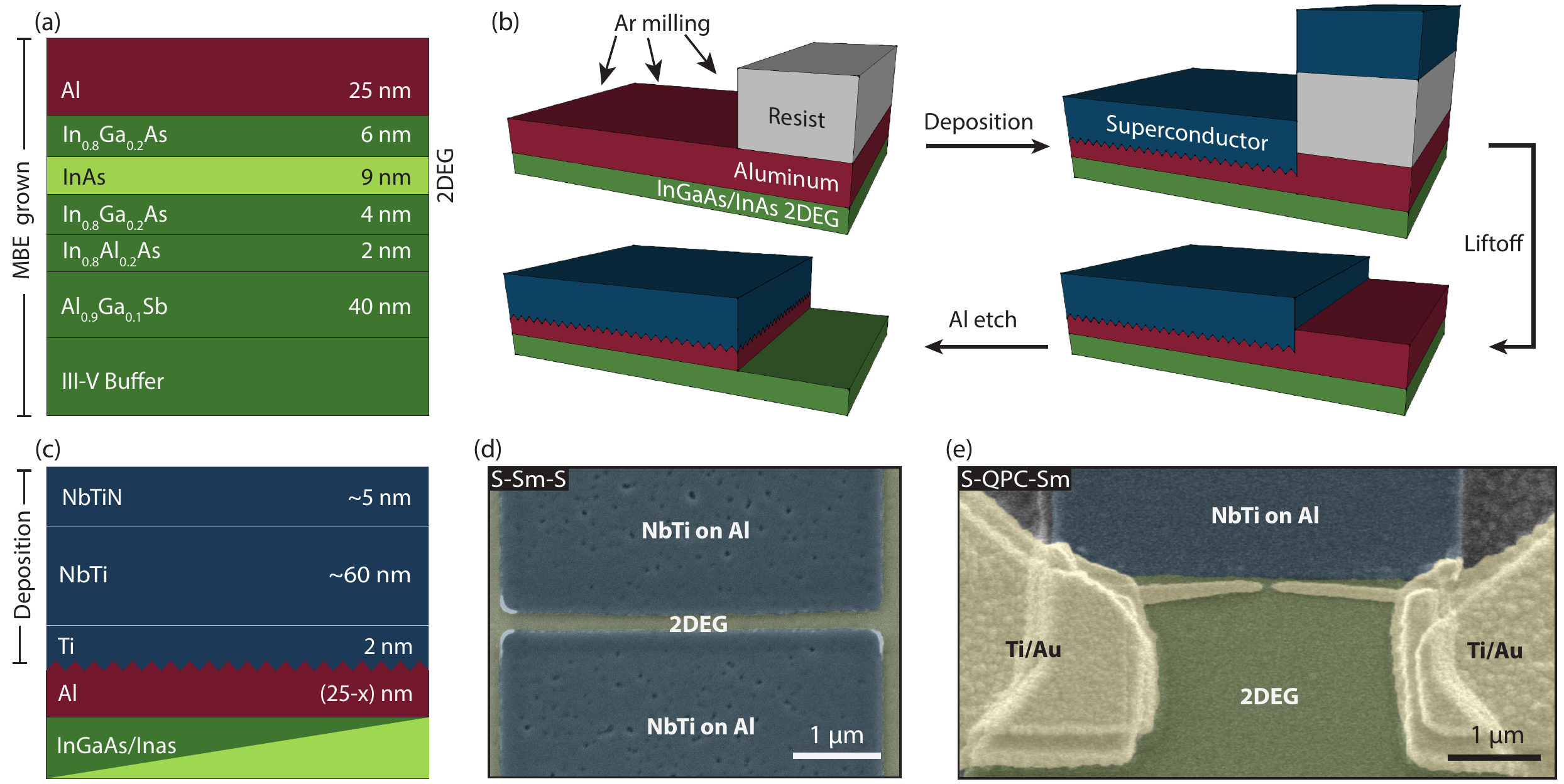}
\caption{(a) MBE-grown InAs/Al hybrid heterostructure. 
(b) Illustration of superconductor deposition and patterning. The native Al oxide is removed in the lithographically defined areas by Ar ion milling, leaving behind a ragged top surface of Al. The NbTi superconductor stack is then deposited without breaking vacuum. After liftoff, the NbTi stack serves as an aligned etch mask for an Al wet etch. 
(c) The NbTi stack includes a Ti layer to promote adhesion and a NbTiN capping to prevent oxidation. 
(d) False-color electron micrograph of an S-Sm-S device after the Al etch similar to the sample measured. 
(e) False-color electron micrograph of the finished S-QPC-Sm device.}
\label{fig1}
\end{figure*}

Intimate coupling between semiconductors (Sm) and superconductors (S) gives rise to novel applications of superconducting electronics \cite{Weinstock1993}, as well as superconducting qubits \cite{Clarke2008} and new topological states of matter \cite{Alicea2011,Leijnse2011}. A critical building block for S/Sm hybrids system is a transparent interface, ensuring high probability of Andreev reflection \cite{BTK,OTBK}. However, obtaining a transparent S/Sm interface has been a technological challenge for decades \cite{Amado2013,Zhong2015,Takayanagi1995}. Recent work has largely resolved the interface problem by realizing epitaxial growth of Al on InAs via molecular beam epitaxy (MBE), both for nanowires \cite{Chang2015,Krogstrup2015} and, more recently, for two-dimensional electron gases (2DEGs) \cite{Shabani2016}, which are better suited for realizing complex, branched devices \cite{Alicea2011}. Focusing on InAs 2DEGs with epitaxial Al, high interface transparency and a hard induced superconductive gap $\Delta^* = 0.18$ meV have been measured both by tunneling spectroscopy via a S-quantum point contact (QPC)-Sm junction \cite{AndreevQPC} and by analysis of multiple Andreev reflection (MAR) in a S-Sm-S junction \cite{MAR}.

Despite its modest superconducting gap, critical temperature and critical magnetic field, Al has been the material of choice for Sm-S epitaxy to date because it is present in conventional III-V MBE systems and is compatible with standard fabrication recipes. {\em In situ} deposition (without breaking vacuum) or direct epitaxial growth of larger gap superconductors has proven challenging and requires dedicated growth systems.  As an alternative, we demonstrate in this work the transfer of large-gap properties of NbTi via the proximity effect through a thin epitaxial Al layer into an InAs 2DEG. 

Starting from an InAs heterostructure with epitaxial Al, we pattern \textit{ex situ} a NbTi-based superconductive layer using standard lithographic techniques. The NbTi layer enhances the Al gap by the proximity effect \cite{Brammertz2001, Cherkez2014}, which in turn results in an enhanced induced gap in the InAs 2DEG. As the processing only involves the topmost Al surface, the high transparency of the epitaxial InAs/Al interface is not affected. 
As discussed below, the induced gap in the InAs 2DEG is found to be more than twice as large as both the induced gap using Al alone \cite{AndreevQPC, MAR} and the gap of the Al itself. The method can be extended to other choices of top-layer superconductor \cite{Brammertz2001}.

We investigate two device geometries, an S-Sm-S Josephson junction and a S-QPC-Sm junction. The S-Sm-S device shows pronounced MAR features, indicating high transparency, and we extract an induced gap, $\Delta^*=0.50$~meV and a critical temperature, $T_{\mathrm{c}}=7.7$~K, with the zero resistance state across the junction persisting up to 3.7 K.
Tunnel spectroscopy in the S-QPC-Sm device yields an induced gap of 0.43 meV and a hardness of the induced gap (measured by a sub-gap conductance suppression in the tunnel regime) comparable to the theoretical limit for S/Sm junctions \cite{Beenakker1992}. Tunnel spectroscopy in a magnetic field reveals a gap closing at a critical in-plane field value of $B_{||,c}\sim 750~\mathrm{mT}$, a value $2.5$ times larger than in similar systems without NbTi \cite{AndreevQPC}.

The InAs heterostructure, grown on an undoped GaSb wafer along the [001] crystallographic direction, is shown in Fig.~\ref{fig1}(a). The active region is similar to previous studies \cite{Shabani2016}, but with a nominal $25~\mathrm{nm}$ epitaxial Al layer, instead of the 10 nm Al layer used previously. Thicker Al allowed this layer to be thinned during fabrication without risking etching down to the interface, as discussed below.
Transport measurements in a Hall bar with the Al removed were used to extract a mobility of $10,000~\mathrm{cm^2/Vs}$ at a density of $1\times 10^{12}~\mathrm{cm}^{-2}$.

Critical fabrication steps are outlined in Fig.~\ref{fig1}b, with full details given in the Supplemental Material~\cite{Supplement}. First, areas for NbTi deposition are patterned using electron-beam lithography. Inside the deposition chamber, the native oxide on Al is removed using Kaufman Ar milling, followed immediately by evaporating Ti and sputtering NbTi/NbTiN without breaking vacuum, as shown in Fig.~\ref{fig1}(c). The Ti bottom layer promotes adhesion and the NbTiN top layer prevents subsequent oxidation. Following liftoff, the patterned NbTi forms a self-aligned mask for a selective Al etch. Fabrication steps not shown in Fig.~\ref{fig1}(b) include mesa etching, deposition of $50~\mathrm{nm}$ of $\mathrm{HfO_2}$ by atomic layer deposition, and evaporation of Ti/Au top gates. 

Figure~\ref{fig1}(d) shows an S-Sm-S device similar to the one measured. The measured device had a separation between the superconducting banks of $150~\mathrm{nm}$ and a width of $4.3~\mathrm{\mu m}$ \footnote{Image is taken before etching, ALD and gate deposition.}. Figure~\ref{fig1}(e) shows the measured S-QPC-Sm device. 
All measurements were carried out using standard AC lock-in techniques in a dilution refrigerator with base temperature of $35~\mathrm{mK}$. The S-Sm-S device was measured with a $5~\mathrm{nA}$ current bias; the S-QPC-Sm device was measured with a $10~\mathrm{\mu eV}$ voltage bias. 

\begin{figure}
\includegraphics[width=0.48\textwidth]{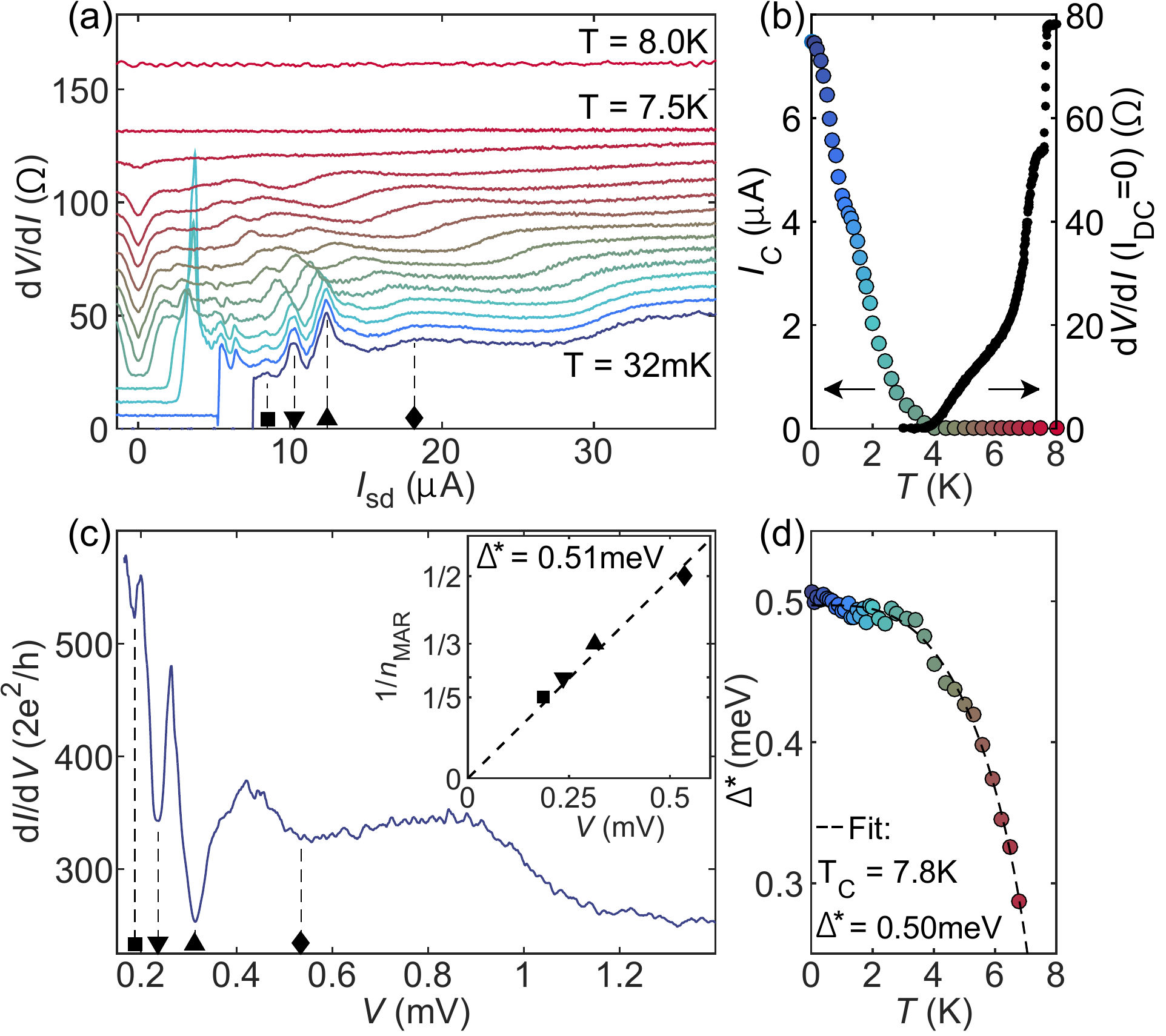}
\caption{Characterization of a S-Sm-S device similar to that shown in Fig.~\ref{fig1}(d). (a) Differential resistance as a function of temperature, showing harmonic structures due to MAR resonances. Traces are successively offset by 6 $\Omega$. Peaks in resistance corresponding to different order of MAR resonances are highlighted by symbols: $\blacklozenge$: 2nd, $\blacktriangle$: 3rd, $\blacktriangledown$: 4th and $\blacksquare$: 5th. (b) Critical current (left axis) and zero bias resistance (right axis) as a function of temperature. $I_\mathrm{c}$ is extracted from (a) and the zero bias resistance was measured while increasing temperature, here binned in steps of 0.01 K. (c) Differential conductance vs voltage at base temperature showing dips/peaks arising from MAR. 2nd to 5th order MAR are highlighted as in (a). Inset shows how the induced gap, $\Delta^*$, is extracted from linear fit to the MAR formula $eV = 2\Delta^{*}/n_{\mathrm{MAR}}$. (d) Temperature dependence $\Delta^*$ extracted as shown in the inset of (c), is fitted by Eq. \ref{eq1} with $T_c$ and $\Delta^*$ as parameters. Colors are used to enable comparison of data between sub-figures.}
\label{fig2}
\end{figure}

The induced gap  under the superconducting leads \cite{OTBK,Flensberg1988} and the interface transparency \cite{Chrestin1997} can be extracted from MAR measurements in the S-Sm-S junction. As discussed in Ref.~\cite{MAR}, a characteristic feature of InAs/epitaxial Al Josephson junction is that sub-gap MAR features appear as peaks in resistance rather than in conductance, a consequence of the high S/Sm interface transparency. Differential resistance $dV/dI$ as a function of DC bias $I_{\mathrm{DC}}$ at various temperatures is shown in Fig.~\ref{fig2}(a). At base temperature, the critical current, $I_\mathrm{C}=7.5~\mathrm{\mu A}$, with normal-state resistance $R_{\mathrm{N}}=53~\mathrm{\Omega}$ reached at $I_{\mathrm{DC}}=45~\mathrm{\mu A}$. The $I_{\mathrm{C}}R_{\mathrm{N}}$ product was $0.40~\mathrm{meV}$ and critical current density was $1.74~\mathrm{\mu A/\mu m}$, both considerably larger than that measured S-Sm-S junctions with epitaxial Al alone \cite{MAR}. 

As seen in Fig.~\ref{fig2}(b) the junction remains in the zero-resistance state up to $3.7~\mathrm{K}$, above which resistance increases up to a plateau at $53~\Omega$, which we interpret as the normal-state resistance of the junction. The sharp transition at $7.7~\mathrm{K}$ is associated with the critical temperature of the NbTi contacts.
The differential resistance displays pronounced MAR features, appearing as sharp peaks in resistance [or dips in conductance, see Fig.~\ref{fig2}(c)]. 

From the MAR relation $eV = 2\Delta^{*}/n_{\mathrm{MAR}}$, a linear fit of the inverse MAR index $1/n_{\mathrm{MAR}}$ as a function of the voltage bias $V$, [Fig.~2(c) inset] yields a base temperature induced gap $\Delta^*=0.51~\mathrm{meV}$. Repeating the procedure over a range of temperatures yields $\Delta^*(T)$, shown in Fig.~2(d) along with a fit to the BCS form \cite{Tinkham}, 
\begin{equation}
\Delta^* (T) = \Delta^*(0)\tanh\left(1.74\sqrt{\frac{T_c}{T}}-1\right),
\label{eq1}
\end{equation}
giving $\Delta^*(0)=0.50~\mathrm{meV}$ and $T_\mathrm{c}=7.8~\mathrm{K}$ as fit parameters. Experimental data and the BCS fit are in good agreement.

\begin{figure}
	\centering
	\includegraphics[width=0.48\textwidth]{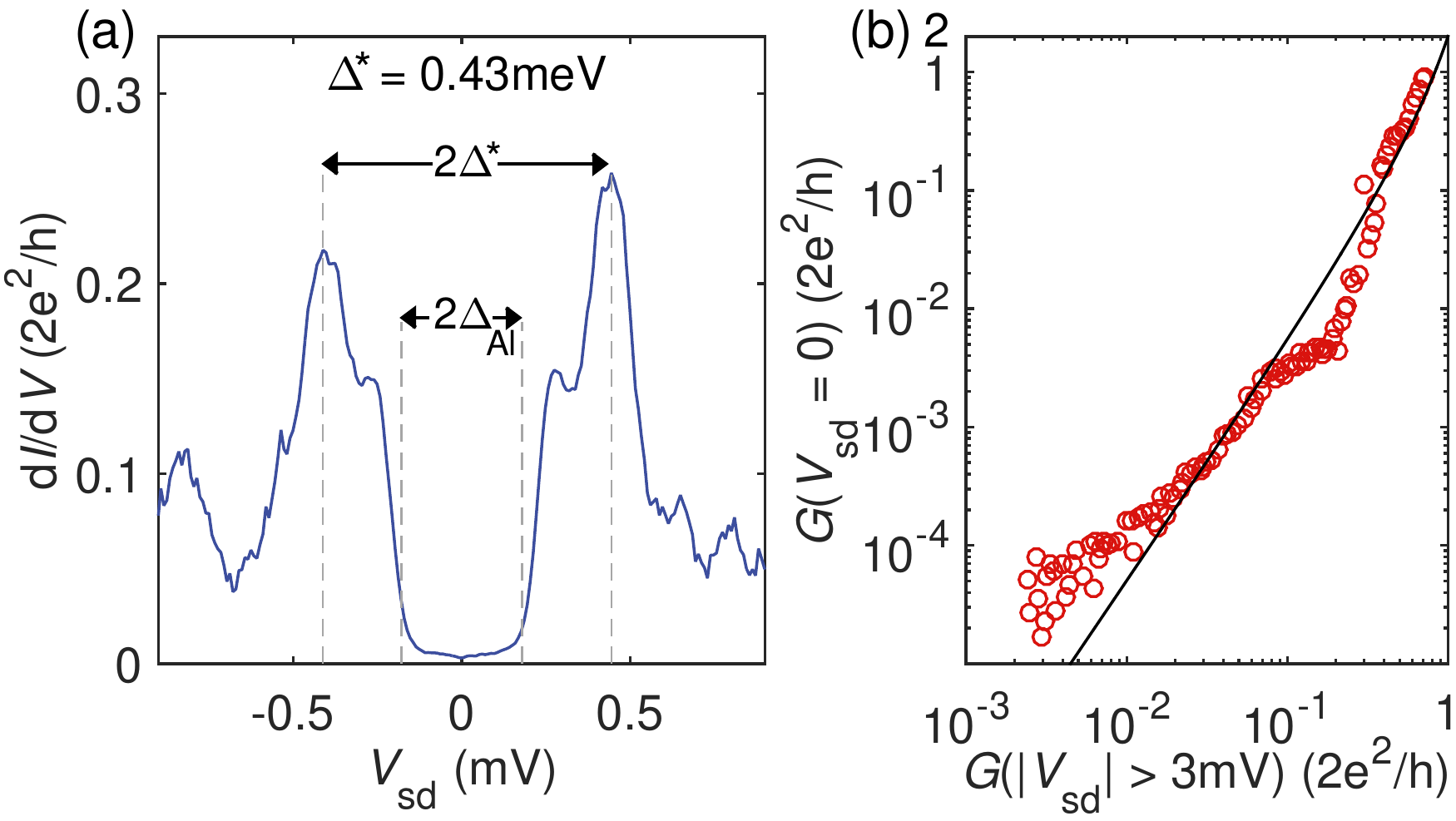}
	\caption{
	Characterization of the S-QPC-Sm device shown in Fig.~\ref{fig1}(e). (a) Differential conductance d$I$/d$V$ as a function of the voltage drop across the QPC $V_{\mathrm{sd}}$. Data is measured with the S-QPC-Sm device in the tunnel regime. An induced gap, $\Delta^*$, is extracted as half the peak-to-peak distance. For comparison the gap of bulk Al $\Delta_{\mathrm{Al}}=0.18~\mathrm{meV}$ is also displayed.  (b) Zero-bias conductance, $G(V_{\mathrm{sd}}=0)$ versus an averaged of conductance at high finite bias $G(|V_{\mathrm{sd}}|>3.0~\mathrm{mV})$.
		Values are extracted from a bias-gate-2D map of conductance, shown in the appendix and are slightly binned. Black curve is a theoretical prediction for perfectly transparent junctions, eq. \ref{eq2}, with no free parameters.
		}
	\label{fig3}	
\end{figure}

A requirement for the use of S/Sm devices for certain applications, including topological quantum computing, is the absence of sub-gap states, reflected in a small sub-gap conductance. Using the approach of Ref.~\cite{Chang2015,AndreevQPC}, we measured electron tunneling near the S/Sm interface using a gate-defined QPC fabricated on top of an ALD oxide. A false colored SEM of the device is shown on Fig.~\ref{fig1}(e).

Figure~\ref{fig3}(a) shows the differential conductance $dI/dV$ as a function of source-drain bias $V_{\mathrm{SD}}$ where the gates are energized to set the QPC in the tunneling regime. In this case, the differential conductance maps the local density of states, allowing the induced gap $\Delta^*$ at the position of the QPC to be directly measured. Defining $2\Delta^*$ as the peak to peak separation gives $\Delta^*=0.43~\mathrm{meV}$, as shown in Fig.~\ref{fig3}(a), similar to the value measured via MAR. In addition to the large energy gap---a factor $\sim 2.3$ larger than the Al-only case \cite{AndreevQPC})---the hardness of the gap at zero energy is not affected by the additional fabrication. To demonstrate this, we measure similar curves as Fig.~\ref{fig3}(a) for different values of out-of-gap conductance, which result in varying in-gap conductance, and produce the parametric plot of Fig.~\ref{fig3}b (markers). These data are compared to theory of a single mode S/Sm interface \cite{Beenakker1992},
\begin{equation}
G_{\mathrm{ns}} = 2G_0\frac{G_{\mathrm{nn}}^2}{(2G_0-G_{\mathrm{nn}})^2},
\label{eq2}
\end{equation}
with no fit parameters.

Here $G_{\mathrm{ns}}$ is the conductance in the superconducting regime, $G_{\mathrm{nn}}$ is the normal state conductance (measured at high source drain bias) and $G_0=2e^2/h$ is the conductance quantum. The agreement is remarkable up to four order of magnitude, demonstrating our devices operate in the theoretical limit of low in-gap conductance. The complete data set is presented in the Supplemental Material~\cite{Supplement}.

\begin{figure}
	\centering
	\includegraphics[width=0.48\textwidth]{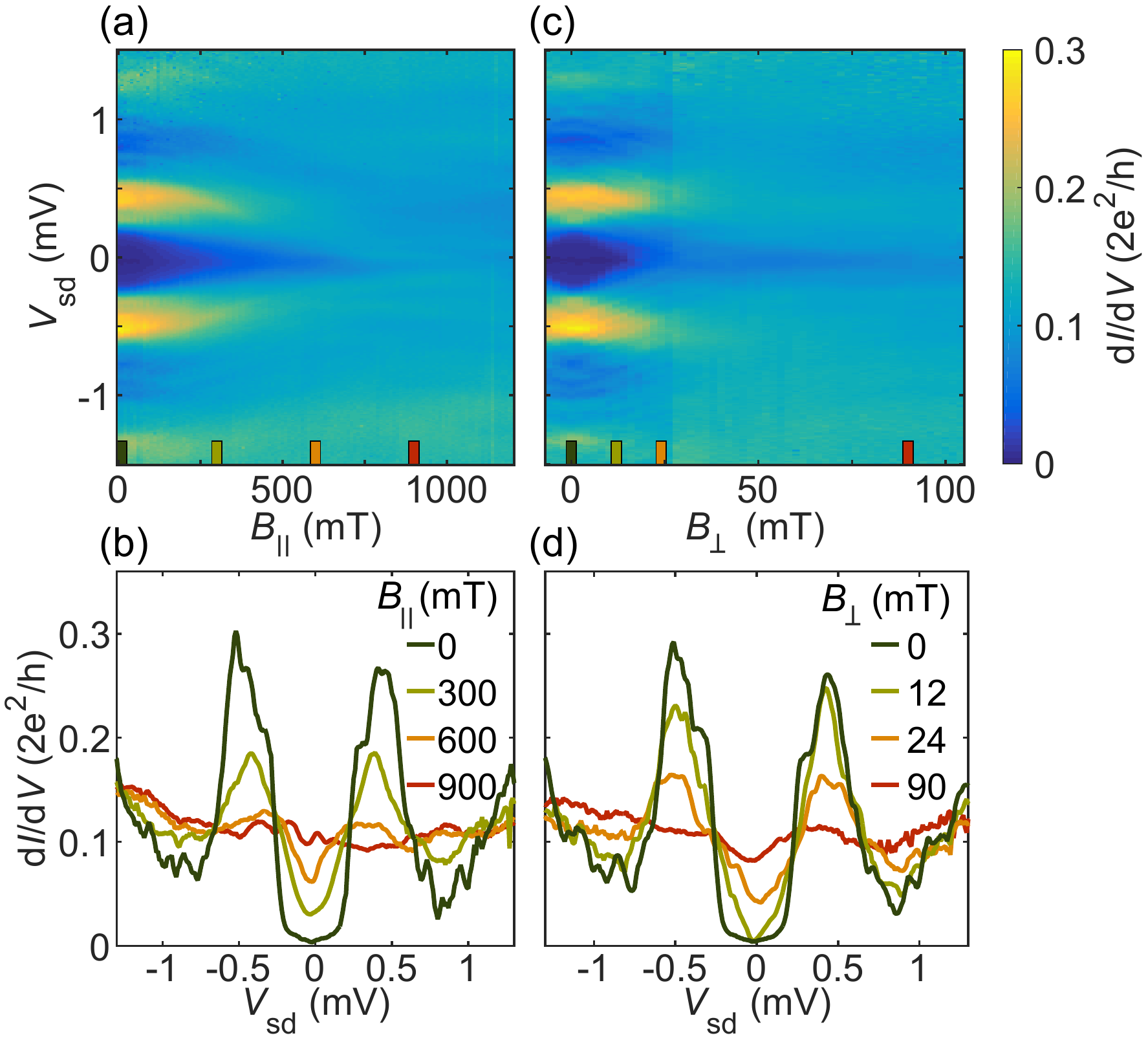}
	\caption{
		Magnetic field dependence of the S-QPC-Sm device of Fig.~\ref{fig1}(e). (a) Differential conductance d$I$/d$V$ as a function of source-drain voltage $V_\mathrm{sd}$ and parallel field $B_{||}$, perpendicular to the current. Four colored lines show the $B_{||}$ values of the line cuts in (b). 
		(c) d$I$/d$V$ as a function of $V_\mathrm{sd}$ and out-of-plane field $B_{\perp}$. Four colored lines show the $B_{\perp}$ values of the line cuts in (d)}
	\label{fig4}
\end{figure}

In order to drive the system in the topological regime, a necessary condition is that the Zeeman splitting in InAs, $g_{\mathrm{Sm}}\mu_{\mathrm{B}}B$, exceeds $\Delta^*$ while the parent superconductor remains gapped. Previous measurements on InAs/Al heterostructures showed a gap closing at in-plane fields $B_{||}\sim 300~\mathrm{mT}$, compatible with $g_{\mathrm{Sm}}\sim 10$, a reasonable value for InAs \cite{AndreevQPC}. Similar measurements presented in Fig.~\ref{fig4}(a,b) indicate a gap closing for in-plane magnetic fields of $B_{||,\mathrm{c}}\sim 750~\mathrm{mT}$, consistent with an induced gap $2.3$ times larger than experiments with epitaxial Al. We note that gap closing is not linked to the quenching of superconductivity in the parent superconductor as NbTi can sustain much larger fields than those used here. A different situation is observed when the field is applied out-of-plane, as shown in Fig.~\ref{fig4}(c,d). In this case the most prominent effect is not a gap closing (the $\pm\Delta^*$ peaks do not approach zero), but rather a gap softening above $100~\mathrm{mT}$ that makes the gap indistinguishable from the background. We interpret the softening as due to dephasing of Andreev pairs in the presence of vortices penetrating the S/Sm stack in a large out-of-plane fields \cite{Tinkham}.

The possibility to locally control the induced gap (and critical field) in the Sm by combining regions with only epitaxial Al or with NbTi/Al stacks allows the realization of complex devices required for future studies of topological states of matter \cite{Aasen2016}. For example, NbTi could be used to realize superconductive leads that persist in the trivial regime while one dimensional devices, proximatized by epitaxial Al only, undergo the topological transition.

In summary, we have demonstrated a method for obtaining a S/Sm heterostructure with high interface transparency and a large, controllable induced gap. The processing is based on MBE grown InAs/Al heterostructures and {\em ex-situ} deposition of a large-gap superconductor. The technique does not compromise the epitaxial interface and so should be compatible with a variety of materials and processing technologies.  Our results suggest a path toward semiconductor-superconductor electronics, both conventional and topological, operating in the temperature range of liquid helium or pulse tube coolers.

\begin{acknowledgments}
We thank S. Upadhyay for useful discussions on fabrication. Research supported by Microsoft Station Q, the Danish National Research Foundation, the Villum Foundation, and the European Commission through a Marie Curie Fellowship. 
\end{acknowledgments}

\bibliography{Publication}

\section{Supplemental Material}
This supplemental Material Section describes the experimental procedures used to fabricate the devices presented in the main text and presents additional electrical measurements

\setcounter{equation}{0}
\renewcommand{\theequation}{S.\arabic{equation}}
\setcounter{figure}{0}
\renewcommand{\thefigure}{S.\arabic{figure}} 

\section{Fabrication techniques}
Here we present detailed information on the fabrication of the reported devices. All patterning was done by e-beam lithography and unless otherwise stated, standard PMMA resist was used. To reduce exposure time, all designs were divided into inner structures with small beam current and outer structures with larger beam current. 

To enable alignment of consecutive exposures, alignment marks are made from a Ti/Au (5/100nm) deposition. The marks are placed near the edge, all around the 2.5$\times$5~mm$^2$ chip.

\subsection{Argon Milling and NbTi deposition}

Starting from the blank chip with alignment markers, we spin coat an MMA/CZAR resist bilayer. The choice of this resist stack is particularly important for a successful Ar etching and deposition step. In particular, the bottom MMA layer provides a sizable undercut to facilitate the lift-off of closely spaced contacts while the CZAR layer can sustain prolonged Ar milling times.

Argon Milling is used to remove the native oxide that forms on the epitaxial Al upon exposure to air, allowing the NbTi to directly contact the metallic Al. For this step we used a Kaufman ion source installed in the superconductor deposition chamber. The etching was performed with a beam voltage of 600~V, an acceleration voltage of 120~V, an Ar-flow of 30~sccm at 1~mTorr pressure and with a rotating sample plate. The desired etching depth in the epitaxial Al is between $10$ and $20~\mathrm{nm}$.

The etching rate of Kaufman source can fluctuate over long periods of time, requiring a fine tuning of the etching time prior to each run. In order to reduce the consumption of epitaxial material, two distinct etching rate calibrations was performed on Si chips with thermally grown SiO$_2$ layer. We found that etching 24.5~nm of SiO$_2$ resulted in an optimal etch depth in the epitaxial Al layer. The etching depth was measured with an optical profilometer for epitaxial Al and a spectral reflectometer for SiO$_2$.
For every etching session, the Kaufman filament was heated by a $10$~minutes milling of an empty sample plate. The sample was subsequently loaded in the deposition chamber and two more minutes of milling were performed on a closed sample shutter before allowing the Ar ions to reach the sample. The samples presented in the Main Text were etched 3~min and 31~sec.

Immediately after Ar etching the superconductive Ti/NbTi/NbTiN stack is deposited in the same chamber. First 2~nm of Ti are e-beam evaporated to ensure good contact between NbTi and Al. Second, 60~nm of NbTi are deposited from a Nb$_{2/3}$Ti$_{1/3}$ sputter target with a beam power of 200~W in a 4~mTorr pressure with an Argon flow rate of 50~sccm, resulting in a deposition rate of $10~\mathrm{nm~min^{-1}}$. When the desired NbTi deposition is terminated, the sample shutter is closed. Third, a N$_2$ flow of 6~sccm is let into the chamber and the shutter is opened again after 30~sec to deposit 5~ nm of NbTiN. The sample is kept rotating during the entire deposition to ensure uniformity.

After deposition, the resist bilayer is lifted-off by immersion of the sample in dioxolane. Sputtering deposition on a undercut resist results in prominent sidewalls, high and narrow structures that can cause several problems during the remaining fabrication steps. Most sidewalls were removed by sonication during liftoff of the sputtered material.

The epitaxial Al wafers do not require specifically designed bonding pads, as low resistance ($<500~\mathrm{\Omega}$) contacts can generally be obtained by directly bonding on the epitaxial Al. In the present case, however, the epitaxial Al covering the surface (and not protected by NbTi) will be removed in a subsequent step. For this reason it is important to deposit NbTi also on the bonding pads of the mesa structure.

\subsection{Mesa Etching and Al etching}
A new resist-pattern is defined for mesa etching on standard PMMA. After chemical development with MIBK:IPA 1:3, the chip is plasma ashed for 60~sec to remove possible resist leftovers. The etching is performed in two steps. First the epitaxial Al, that covers the entire surface of the wafer, must be removed. Second, the III/V semiconductor is etched to isolate different devices on the same chip.
The epitaxial Al is removed by a 12~sec etching in 50~$^\circ$C Transene Aluminum Etchant type D. The process is terminated first with 30~s stirring in 50~$^\circ$C DI water and then 30~s stirring in room temperature DI water. 

After blow drying the chip with nitrogen, the III-V is immediately etched by a prepared room temperature H$_2$O:C$_6$H$_8$O$_7$:H$_3$PO$_4$:H$_2$O$_2$ (220:55:3:3) solution for 330~s, resulting in an etching dept of about 600 ~nm. The ethcing is stopped with stirring in room temperature DI water. The resist is lifted off with dioxolane, followed by washes in acetone and isopropanol. At this point the NbTi forms a self aligned mask that can be used for etching the epitaxial Al covering the mesas, so no further electron beam lithography step is needed. The Al etching step previously described is then repeated, but with no resist mask.

\subsection{Atomic Layer Deposition of HfO$_2$ and gate deposition}
Immediately after removing the unwanted epitaxial Al, the chip is transferred in an atomic layer deposition chamber for the growth of 50~nm HfO$_2$ as a gate insulator. To minimize the exposure of the uncovered III-V material to oxygen, a constant flow of 20~sccm of N$_2$ is maintained in the chamber at any stage, also during pumping down. HfO$_2$ is deposited by 500 cycles of Tetrakis(dimethylamido)hafnium pulse and 60 sec waiting time and water pulse and 60 sec waiting time preheated at 90~$^\circ$C.

Top gate deposition is done in two steps, one for the fine features and the other for larger elements such as bonding pads. In Fig.~1(e) of the Main Text it is possible to distinguish the two depositions from their different metal height and surface roughness. The fine features are defined in a single PMMA layer by evaporation of 5~nm Ti/30~nm Au and lift-off. The larger features are defined in a MMA/PMMA bilayer and require the evaportation of 50~nm Ti/700~nm Au. In both cases the chips are plasma ashed for 60~s after development to remove eventual resist leftovers.

\section*{S-Sm-S in-plane field dependence}

\begin{figure}
\includegraphics[width=\columnwidth]{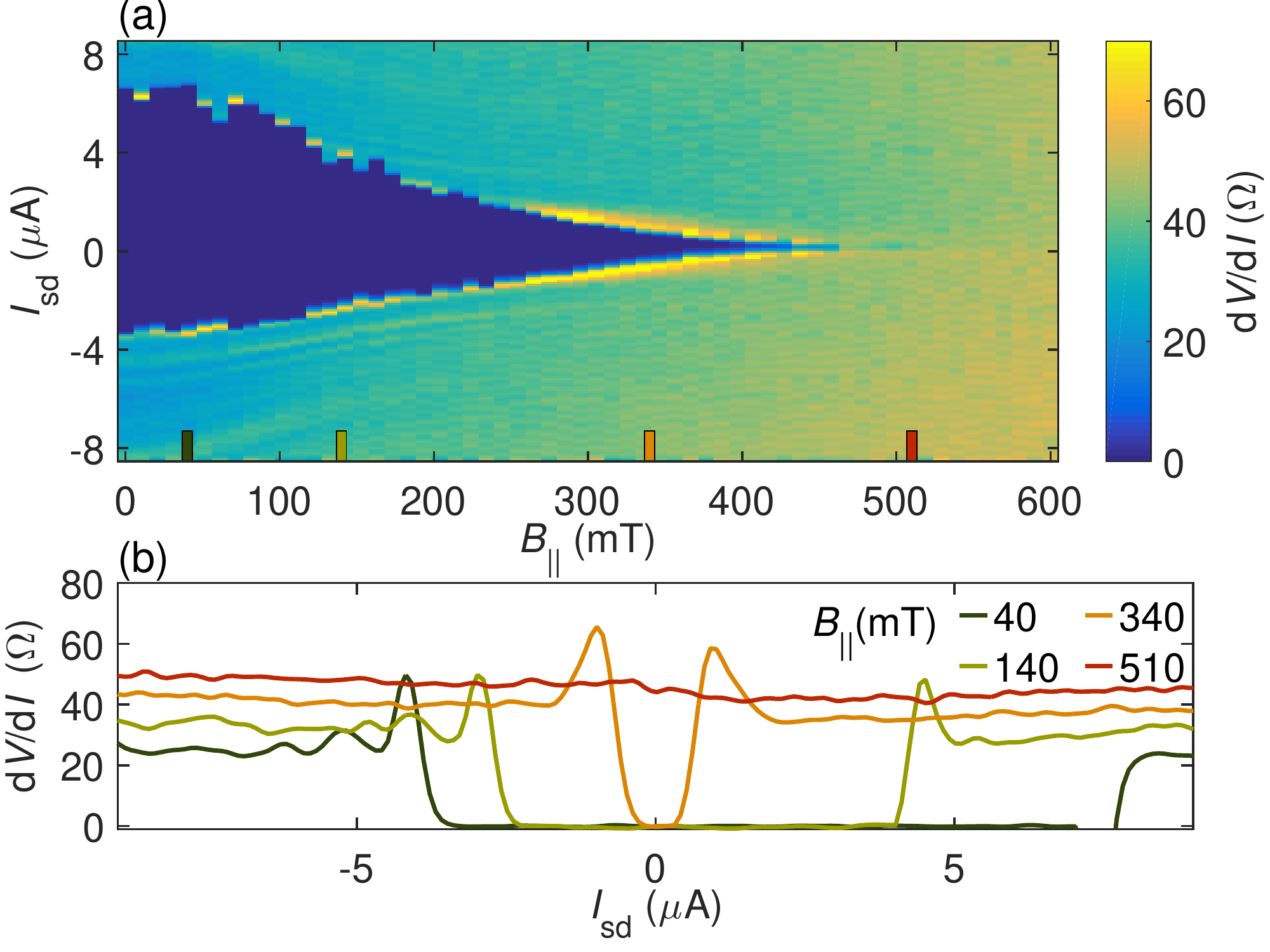}
\caption{In-plane field dependence of the S-Sm-S. (a) Differential resistance $dV/dI$ as a function of source-drain current $I_{\mathrm{sd}}$ and in-plane field $B_{||}$, applied perpendicular to the current flow. Colored squares show $B_{||}$-values at which line cuts in (b) are taken. These line cuts are not offset.}
\label{figs1}
\end{figure}

An in-plane field, perpendicular to the current direction, is applied on the S-Sm-S junction while measuring critical current $I_{\mathrm{C}}$, see Fig. S\ref{figs1}. The vector-magnet was aligned before the measurement. Still $I_{\mathrm{C}}$ is fluctuating at low fields, indicating flux jumps. These stabilizes at higher fields until the super-current dies out at $\sim$500 mT. 
The sweeps are taken going from negative to positive, causing heating effects in the negative current region giving rise to the visible asymmetry of $I_{\mathrm{C}}$.
\newline

\section{QPC Conductance}
\begin{figure}
\includegraphics[width=0.85\columnwidth]{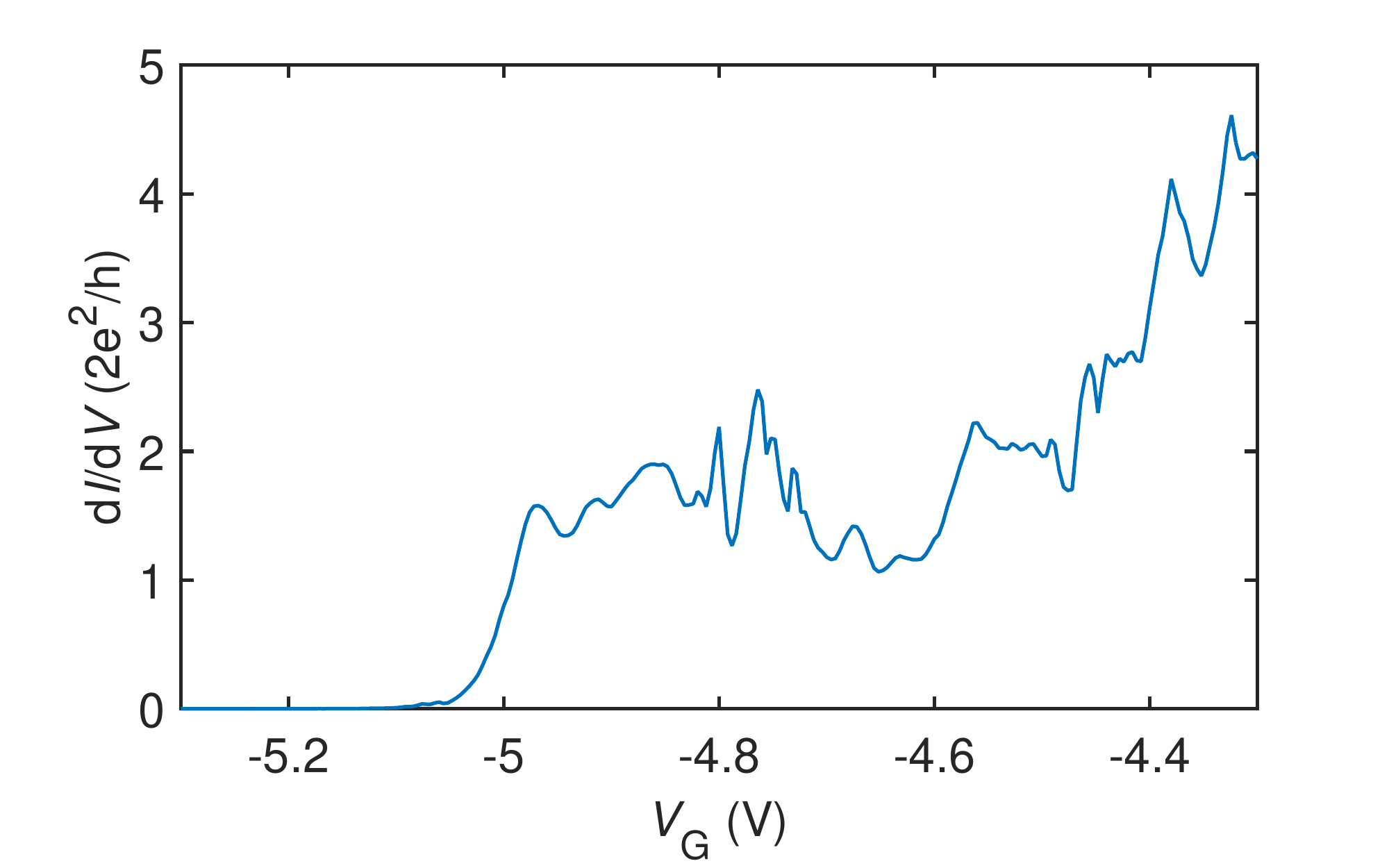}
\caption{Differential conductance $dI/dV$ as a function of gate voltage $V_{\mathrm{G}}$, at zero bias, close to the pinch-off of the QPC.}
\label{figs2}
\end{figure}

Beenakker predicted in 1992 \cite{Beenakker1992} that an Andreev-enhanced QPC, like the one reported in this work, should have $4e^2/h$ steps in conductance after pinch-off. This was observed in an InAs 2DEG with epitaxial Aluminum S-QPC-Sm junction \cite{AndreevQPC}. Despite the reported S-QPC-Sm does not show clear step features, presumably for the lower electron mobility of the wafer used in this work, the first plateau fluctuates around $4e^2/h$ rather than $2e^2/h$, see figure. S\ref{figs2}.

\subsection{Conductance in the tunneling regime}
\begin{figure*}
\includegraphics[width=2\columnwidth]{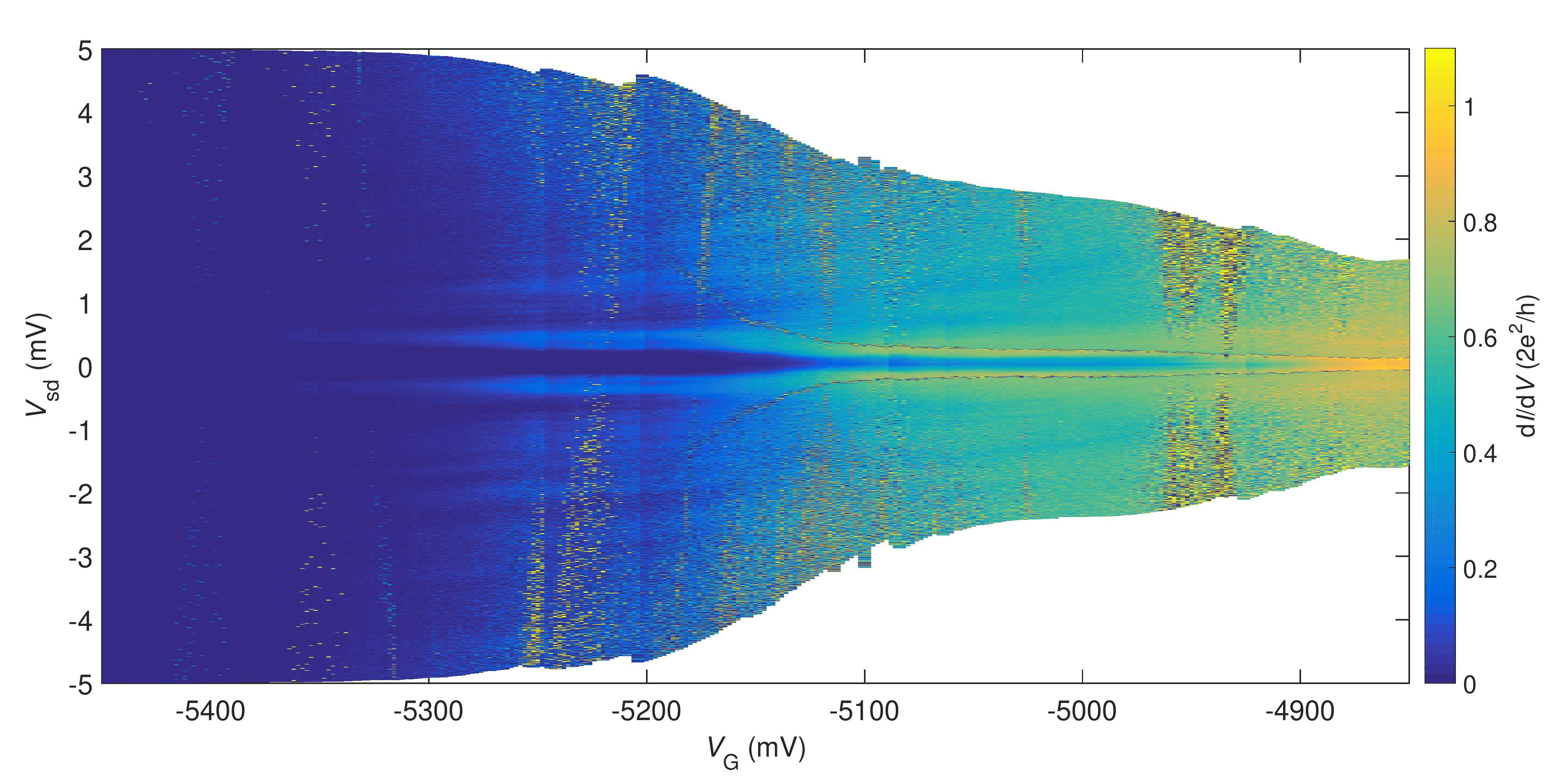}
\caption{Numerically differentiated current $dI/dV$ vs source-drain voltage $V_{\mathrm{sd}}$ for gate voltages $V_{\mathrm{G}}$ close to pinch-off with the QPC.}
\label{figs3}
\end{figure*}

The parametric plot presented in Fig.~3(b) of the Main Text shows the QPC zero bias conductance as a function of the QPC high-bias conductance (which coincides with the normal state conductance up to experimental errors). The parametric plot was obtained from the measurement presented in Fig.~\ref{figs3}, showing the QPC conductance as a function of bias $V_{\mathrm{sd}}$ and gate voltage $V_{\mathrm{G}}$ close to pinch-off.

To accurately measure the sample conductance in the very low transmission regime, we used DC techniques only in a two-terminal configuration. A line resistance $R_\mathrm{Line}=11.8\mathrm{k\Omega}$ was determined in a four-terminal measurement at $V_\mathrm{G}=-4.85~\mathrm{V}$. Figure~Fig.~\ref{figs3} is obtained by numerical differentiation of the measured DC current as a function of $V_{\mathrm{SD}}$. The voltage dropping on the QPC was calculated as $V_\mathrm{eff}=V_\mathrm{sd}-I_\mathrm{DC}R_\mathrm{Line}$. The noise visible in Fig.~S\ref{figs3} is presumably due to a combination of the differentiation method and intrinsic noise of the device, also noticed in AC measurements. 

\end{document}